\documentclass[a4paper]{bio}\usepackage[]{graphicx}\usepackage[]{color}
\makeatletter\let\ps@plain\ps@myheadings\makeatother 
\makeatletter
\def\maxwidth{ %
  \ifdim\Gin@nat@width>\linewidth
    \linewidth
  \else
    \Gin@nat@width
  \fi
}
\makeatother

\usepackage{fixltx2e}         
\usepackage[utf8]{inputenc}   
\usepackage{url}

\usepackage{amsmath, amssymb, amsfonts} 
\usepackage{mathtools}        
\usepackage{bm}               
\usepackage{dsfont}           

\usepackage{booktabs}         
\usepackage{subcaption}       
\newcommand{\subfloat}[2][need a sub-caption]{\subcaptionbox{#1}{#2}} 
\let\code=\texttt
\let\proglang=\textsf
\newcommand{\pkg}[1]{\code{#1}}

\newcommand{\normalize}[1]{\lfloor#1\rfloor}

\newcommand{\abk}[1]{\mbox{#1}.\ }
\newcommand{\eg}{\abk{e.g}}
\newcommand{\ie}{\abk{i.e}}
\newcommand{\cf}{\abk{cf}}

\title{Incorporating social contact data in spatio-\\temporal models for infectious disease spread}
\author{\uppercase{Sebastian Meyer}$^\ast$\\[4pt]
\textit{
  Institute of Medical Informatics, Biometry, and Epidemiology, 
  Friedrich-Alexander-Universit\"at Erlangen-N\"urnberg,
  Waldstra{\ss}e 6,
  DE-91054 Erlangen,
  Germany
}\\[2pt]
{seb.meyer@fau.de}\\[8pt]
\uppercase{Leonhard Held}\\[4pt]
\textit{
  Epidemiology, Biostatistics and Prevention Institute, 
  University of Zurich,
  Hirschengraben 84,
  CH-8001 Z\"urich,
  Switzerland
}
}
\IfFileExists{upquote.sty}{\usepackage{upquote}}{}
\begin{document}


\maketitle
\footnotetext{To whom correspondence should be addressed.}

\begin{abstract}
  {
    Routine public health surveillance of notifiable infectious diseases
    gives rise to weekly counts of reported cases---possibly
    stratified by region and/or age group. 
    We investigate how an age-structured social contact matrix
    can be incorporated into a spatio-temporal
    endemic-epidemic model for infectious disease counts.
    To illustrate the approach, we analyze the spread of
    norovirus gastroenteritis over 6 age groups within the 12 districts of Berlin, 2011--2015,
    using contact data from the POLYMOD study.
    The proposed age-structured model outperforms alternative scenarios with
    homogeneous or no mixing between age groups.
    An extended contact model suggests a power transformation of the
    survey-based contact matrix towards more within-group transmission.
  }{ 
    Age-structured contact matrix;
    Areal count time series;
    Endemic-epidemic modelling;
    Infectious disease epidemiology;
    Norovirus gastroenteritis;
    Norwalk virus;
    Spatio-temporal surveillance data.
  }
\end{abstract}

\enlargethispage{\baselineskip}


\section{Introduction}
\label{sec:intro}

The social phenomenon of ``like seeks like'' produces
characteristic contact patterns between subgroups of a population.
If suitably quantified, such social mixing behaviour can inform models for
infectious disease spread 
\citep{read.etal2012}. 
One of the largest social contact surveys to date was conducted as part of the EU-funded POLYMOD project,
recording conversational contacts of 7\,290 individuals in eight European
countries \citep{mossong.etal2008}. Contact patterns were found to be 
similar across the different countries and highly assortative with respect to
age, especially for school children and young adults.

The basic idea behind the combination of social contact data with
epidemic models has been termed the ``social contact hypothesis''
\citep{wallinga.etal2006}: The age-specific numbers of potentially infectious
contacts are proportional to age-specific numbers of social contacts.
For instance, for pathogens transmitted via respiratory droplets,
face-to-face conversation and/or physical contact are frequently used as proxy measures
for exposure. 
Many studies have now made use of the POLYMOD contact data
\citep{rohani.etal2010, goeyvaerts.etal2010, goeyvaerts.etal2015, birrell.etal2011, baguelin.etal2013},
but none of them accounts for the spatial characteristics of disease spread.
The distance of social contacts from the home location
of each participant has only recently been investigated by
\citet{read.etal2014}. Their finding that ``most were within a
kilometre of the participant's home, while some occurred further than
500 km away'' reflects the power-law distance decay
 of social interaction as determined by human
travel behaviour \citep{brockmann.etal2006}. 
\citet{meyer.held2013} found such a power law to translate to the spatial spread of infectious diseases.

The purpose of this paper is to combine the social and spatial determinants
of infectious disease spread in a multivariate time-series model for public health
surveillance data. For notifiable diseases,
such data are routinely available as weekly counts of reported
cases by administrative district and
further stratified by age group or gender. 
Social contact matrices reflect the amount of mixing between these strata.
Our focus is on
age-structured models, but the methods equivalently apply to
other or multiple strata.
We investigate if a (possibly adjusted) contact matrix captures
disease spread better than simple assumptions of homogeneous or no mixing
between the subgroups.
The approach also allows us to estimate how much disease incidence in
each group can be linked to previous cases in their own and in other
groups---while adjusting for the spatial pattern of disease spread.

This paper is organized as follows.
Section~\ref{sec:data} introduces our case study on 
norovirus gastroenteritis,
including contact data from the POLYMOD study.
Section~\ref{sec:methods} outlines the spatio-temporal
modelling framework and describes how to incorporate additional stratification 
with a contact matrix.
Section~\ref{sec:results} shows results of the case study 
and Section~\ref{sec:discussion} concludes the paper with a discussion.
The supplementary material contains additional figures, an animation of the data,
as well as the \proglang{R}~source package \pkg{hhh4contacts} with
the data and code to reproduce the presented analysis
(run \code{demo("hhh4contacts")} after installing and loading the package).

\section{Case study: Norovirus gastroenteritis in Berlin, 2011--2015}
\label{sec:data}

Most of the aforementioned studies relate contact patterns to the
spread of influenza, whereas here we  investigate the occurrence of norovirus-associated
acute gastroenteritis. Both diseases are highly infectious, have a similar
temporal pattern, and similar mortality in elderly persons
\citep{asten.etal2012}. However, in contrast to influenza, vaccines against
noroviruses have yet to be developed \citep{pringle.etal2015}.
Absence of vaccination
simplifies the analysis of infectious disease occurrence since vaccination
coverage---potentially varying across age groups, regions and over time---needs not to be
taken into account.

\subsection{Epidemiology of norovirus gastroenteritis}

Norovirus-associated acute
gastroenteritis is characterized by ``sudden onset of vomiting, diarrhea,
and abdominal cramps lasting 2--3 days''
\citep{pringle.etal2015}. \citet{o'dea.etal2014} estimate an average symptomatic period of 3.35
days from outbreaks in hospitals and long-term care facilities,
where vulnerable individuals live closely together and norovirus outbreaks most
commonly occur. 
Another frequently affected subgroup
are children in daycare centres. Norovirus incidence
peaks during winter, where outbreaks in childcare facilities
were observed to precede those in private households, hospitals, and nursing
homes \citep{norovirus-epidemiology}.


Noroviruses are highly contagious since only few viral particles are needed for
an infection. Being thermally stable and particularly persistent in the
environment \citep{marshall.bruggink2011},
noroviruses can 
also be transmitted indirectly via contaminated surfaces or food.
The serial interval,
\ie the time between onset of symptoms in a primary and a secondary case,
ranges from within a day to more than 1 week with a median of about 3 days
\citep{goetz.etal2001}. 

\subsection{Incidence data}

In Germany, the national public health institute (the Robert Koch Institute,
RKI) provides access to incidence data of notifiable diseases through the
\emph{SurvStat@RKI 2.0} online service (\url{https://survstat.rki.de}).
Since the last revision of the case definition for norovirus gastroenteritis in
2011, only laboratory-confirmed cases are reported to the RKI.
The number of cases to be modelled thus excludes all
asymptomatic cases as well as all those symptomatic cases, who have not found their
way to laboratory testing \citep{gibbons.etal2014}.
It is known that under-reporting of norovirus illness is most pronounced in the
20- to 29-year-old persons and substantially lower in persons aged $<10$ years
and 70 years and over \citep{bernard.etal2014}.
A sensitivity analysis will indicate how under-reporting may affect the
interpretation of our model results.

As to the geographic region of interest, we chose the largest city of Germany,
Berlin, 
which is divided into 12 administrative districts.
This enables the
analysis of disease spread on a smaller spatial scale. Furthermore, a large
underlying population is required for our time-series model
to be a reasonable
approximation of the epidemic process \citep{farrington.etal2003}.

We have downloaded weekly numbers of reported cases of norovirus gastroenteritis
in Berlin 
from \emph{SurvStat@RKI} (as of
the annual report 2015). These counts cover four norovirus seasons, from
2011-W27 to 2015-W26, and are stratified by the 12 city districts and
6 age groups: 0--4, 5--14, 15--24, 25--44, 45--64, and 65+ years of age.
The age groups were condensed from 5-year
intervals to reflect distinct social mixing of pre-school
vs.\ school children, and intergenerational mixing.
Similarly stratified population numbers were obtained from
the Statistical Information System Berlin-Brandenburg \emph{StatIS-BBB}
(\url{https://www.statistik-berlin-brandenburg.de/statis})
at the reference date 31 December 2011,
when Berlin had 3\,501\,872 inhabitants in total.

\begin{figure}

{\centering \includegraphics[width=0.9\linewidth]{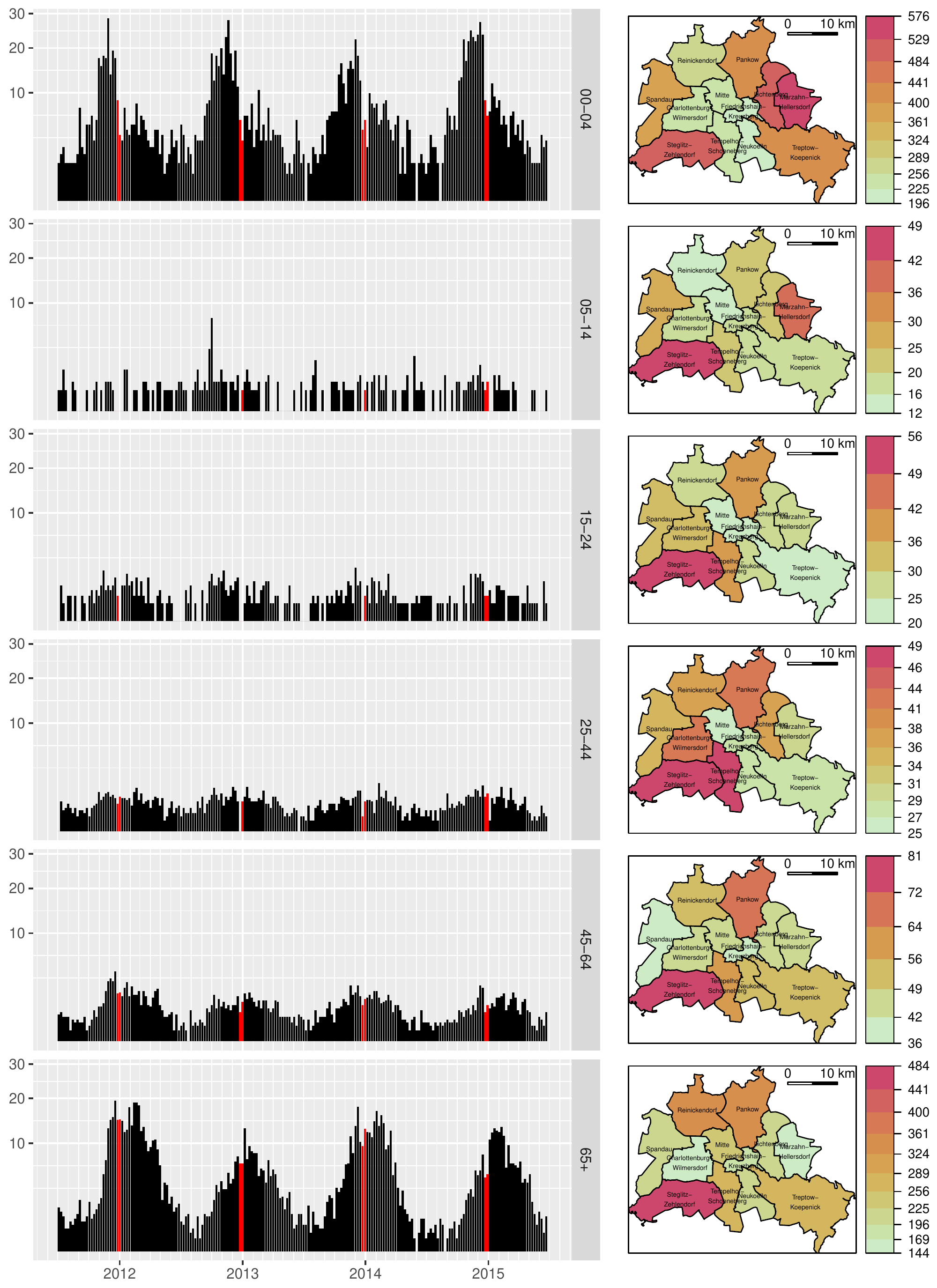} 

}

\caption{Age-stratified time series and maps of norovirus gastroenteritis incidence (per 100\,000 inhabitants) in Berlin, 2011-W27 to 2015-W26. The weekly incidence plots on the left all use the same $\sqrt{}$-scale. The Christmas break in calendar weeks 52 and 1 is highlighted. The group-specific maps on the right show the mean yearly incidence by city district.}\label{fig:incidence_groups}
\end{figure}

Figure~\ref{fig:incidence_groups} (left) shows the weekly norovirus incidence
stratified by age group and aggregated over all city districts. 
The reported incidence is higher in pre-school children and the retired
population than in the other age groups.
The yearly seasonal pattern, with overall counts
ranging from~7 to~214 cases
per week, is approximately constant during the four years 
(supplementary Figure~S1).
The typical bump during the Christmas break could be related to reporting deficiencies
and school closure \citep{hens.etal2009}.
The time series of the~5- to~14-year-old children contains
an outbreak 
caused by contaminated frozen strawberries, which
were delivered almost exclusively 
to schools and childcare facilities \citep{norovirus-outbreak}.
Comparing seasonality between the age groups, the peak incidence in pre-school
children seems to precede the peak in the highest age group.
Our age-structured modelling approach
will help to address the question raised by
\citet{norovirus-epidemiology}, ``whether this reflects a pattern of disease
transmission from young to old in the community''---taking the spatial
aspect of disease spread into account.

How disease incidence varies across the 12 city districts of Berlin is
shown in Figure~\ref{fig:incidence_groups} (right).
The south-western district Steglitz-Zehlendorf tends to be affected more and the
central districts tend to be affected less than the remaining districts.
This pattern is roughly consistent across age groups. An exception are the
two younger age groups, which exhibit a relatively high incidence in Marzahn-Hellersdorf.
District-specific seasonal shifts are not apparent
(supplementary Figure~S2).

Animated, age-stratified maps of the weekly counts
encompass the full information from all three data dimensions. 
Such an animation (supplementary material) may provide
additional insight into the dynamics of disease spread. However,
epidemic models estimated from these data offer a more structured view and
take population heterogeneity directly into account.

\subsection{Contact data}

\begin{figure}

{\centering \includegraphics[width=\textwidth]{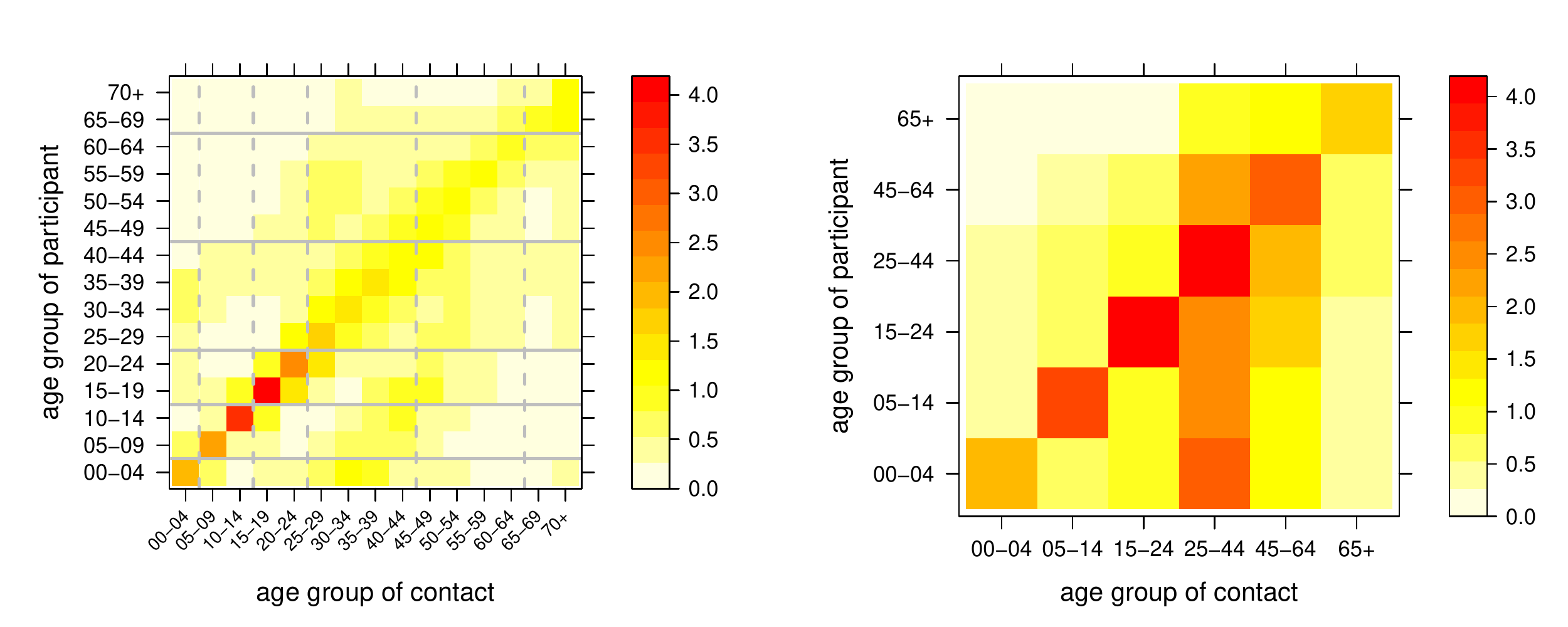} 

}

\caption{Age-structured contact matrix estimated from the German POLYMOD sample using 5-year intervals (left), and aggregated to the age groups of the surveillance data (right). The entries refer to the mean number of contact persons per participant per day.}\label{fig:Cgrouped}
\end{figure}

We use
contact data from the German subset of the POLYMOD study \citep{mossong.etal2008},
where both physical and non-physical (conversational) contacts have been recorded.
We will report results based on all contacts and on physical contacts only.
The age-structured social contact matrix $\bm{C} = (c_{g'g})$ contains
the mean numbers of contact persons in age group~$g$ during one day reported by
a participant in age group~$g'$. Instead of using sample means,
we estimate $\bm{C}$ by the approach of
\citet{wallinga.etal2006}, which accounts for the reciprocal nature of contacts.
Each entry~$c_{g'g}$ is assumed to be the mean of a negative binomial distribution,
under the restriction $c_{g'g} n_{g'} = c_{gg'} n_{g}$, where $n_g$ is Berlin's population in
age group~$g$. We estimate a detailed contact matrix with 5-year intervals,
which we subsequently aggregate to the above 6 age
groups (Figure~\ref{fig:Cgrouped}). Direct estimation of the aggregated contact
matrix leads to similar numbers.

The strong diagonal pattern in the social contact matrix
reflects that people tend to mix with people of the
same age. 
The other prominent pattern is
produced by the contacts between parents and children.
The matrix for physical contacts shows similar patterns (supplementary Figure~S3).
Aggregation of the contact matrix is
done by summing over the contact groups (columns) to be joined and
calculating the weighted average across the corresponding participant
groups (rows), with weights equal to the group sizes.
The aggregated contact matrix is
asymmetric because of the different sizes of the involved age
groups, but reciprocity at the population level still holds.
For the models described in the next
section, only the row-wise distributions will be relevant, \ie the
contact pattern of an infectious participant across the different age
groups.

\section{An age-structured spatio-temporal model for infectious disease counts}
\label{sec:methods}

We review an endemic-epidemic modelling framework
for areal time series of infectious disease counts
\citep[Section~3]{meyer.held2013}, into which we
subsequently incorporate an additional stratification variable
featuring a contact matrix.

\subsection{Spatio-temporal formulation}

Conditionally on past observations, the number of reported infections in
region~$r$ and time period~$t$, $Y_{rt}$, is assumed to follow a negative
binomial distribution with mean~$\mu_{rt}$ and
region-specific overdispersion parameters~$\psi_r$ such that the conditional
variance of $Y_{rt}$ is $\mu_{rt} (1+\psi_r \mu_{rt})$.
The lower bound $\psi_r = 0$ yields the Poisson distribution as a special case,
and a common simplifying assumption is that $\psi_r = \psi$ is shared across regions.
In its most general formulation, the mean $\mu_{rt}$ is
additively decomposed into \emph{endemic} and observation-driven \emph{epidemic} components as
\begin{align} \label{eqn:hhh4}
  \mu_{rt} &= e_{rt} \, \nu_{rt} + \lambda_{rt} \, Y_{r,t-1} +
  \phi_{rt} \sum_{r' \ne r} \normalize{w_{r'r}} \, Y_{r',t-1}\\
\intertext{with log-linear predictors}
\begin{split} \label{eqn:hhh4:predictors}
  \log(\nu_{rt}) &= \alpha_r^{(\nu)} + {\bm{\beta}^{(\nu)}}^\top \bm{z}^{(\nu)}_{rt} \:, \\
  \log(\lambda_{rt}) &= \alpha_r^{(\lambda)} + {\bm{\beta}^{(\lambda)}}^\top \bm{z}^{(\lambda)}_{rt} \:, \\
  \log(\phi_{rt}) &= \alpha_r^{(\phi)} + {\bm{\beta}^{(\phi)}}^\top \bm{z}^{(\phi)}_{rt} \:,
\end{split}
\end{align}
and normalized transmission weights
$\normalize{w_{r'r}} := w_{r'r} / \sum_j w_{r'j}$,
$w_{rr} = 0$.
The regression terms in \eqref{eqn:hhh4:predictors} often include sine-cosine effects of time
to reflect seasonally varying incidence \citep{held.paul2012}, but $\bm{z}_{rt}^{(\cdot)}$
may also involve other explanatory variables, such as vaccination coverage
\citep{herzog.etal2011}.
The first, endemic component in~\eqref{eqn:hhh4} is typically modelled
proportional to a (population) offset $e_{rt}$,
and partially captures infections
not directly linked to observed cases from the previous time period,
\eg due to travelling outside the study region (edge effects).
The 
epidemic component splits up into
autoregressive effects, \ie reproduction of the disease within region~$r$,
and neighbourhood effects, \ie transmission from other regions~$r'$.
It has proven useful to account for population size also in
$\log(\phi_{rt}) = \alpha_r^{(\phi)} + \tau \log(e_{rt})$, 
such that $\tau$ determines how
``attraction'' to a region scales with population size \citep{xia.etal2004}.
Furthermore, transmission weights $w_{r'r}$ reflect the flow of infections
from region~$r'$ to region~$r$.
These weights may be based on additional movement network data
\citep{paul.etal2008,schroedle.etal2012,geilhufe.etal2012}, but may also be
estimated from the data at hand. A suitable parametric model 
is a power-law distance decay $w_{r'r} = o_{r'r}^{-\rho}$
in terms of the adjacency order~$o_{r'r}$ in the
neighbourhood graph of the regions \citep{meyer.held2013}.

Estimating separate dynamics for the reproduction of the disease within a region
on the one hand, and transmission from other regions on the other hand, goes
back to the original model formulation of \citet{held.etal2005}, where only
first-order neighbours have been incorporated. 
The parametric distance weights offer an appealing alternative to reflect
predominant local autoregression in a simpler model with a single epidemic
component: 
\begin{equation} \label{eqn:hhh4_twin}
\mu_{rt} = e_{rt} \, \nu_{rt} + \phi_{rt} \sum_{r'} \normalize{w_{r'r}} \, Y_{r',t-1} \:,
\end{equation}
where the choice $w_{r'r} = (o_{r'r} + 1)^{-\rho}$ gives unit weight to local
transmission ($r'=r$) and then decays as a power law in terms of adjacency order.
With such a power law and the suggested population dependence of $\phi_{rt}$,
the epidemic component of \eqref{eqn:hhh4_twin} constitutes a so-called
gravity model \citep{xia.etal2004,hoehle2016}.
Furthermore, this formulation uses fewer parameters and
extends more naturally to an additional stratification variable.

\subsection{Extension for stratified areal count time series}

Extending the above spatio-temporal
model to fit multivariate time series of counts~$Y_{grt}$ stratified by
(age) group in addition to region, enables us to relax the
simple assumption of homogeneous mixing within each region.
More complex strata such as the interaction of
age group and gender are equally possible and can be subsumed in the
single group index $g$.

We assume that a contact matrix $\bm{C} = (c_{g'g})$ is given,
where each entry $c_{g'g} \ge 0$ quantifies the average number of 
contacts of an individual of group~$g'$ with individuals of group~$g$.
The spatio-temporal model~\eqref{eqn:hhh4_twin}
then extends to a three-dimensional version as
\begin{equation} \label{eqn:hhh4contacts}
  \mu_{grt} = e_{grt} \, \nu_{grt} + \phi_{grt} \sum_{g',r'}
  \normalize{c_{g'g} \, w_{r'r}} \, Y_{g',r',t-1} \:,
\end{equation}
where both the endemic and epidemic predictors may gain
group-specific effects. 
How the counts from the previous period affect the current mean in group~$g$ and
region~$r$ is now determined by a product of contact and spatial weights.
The product ensures that cases from group~$g'$ in region~$r'$ are ignored
if there are no contacts to group~$g$ or if there is no flux of infections
from region~$r'$ to region~$r$. The weights are row-normalized
over all combinations of group and region:
$\sum_{g,r} \normalize{c_{g'g} w_{r'r}} = 1$.
Note that this normalization removes any differences in group-specific overall
contact rates (the row sums of~$\bm{C}$). Our model therefore does not
distinguish between proportionate mixing, where the rows of the contact matrix
only differ by a proportionality factor, and a matrix with identical rows.
The weighted sum of past cases transmitted
to group~$g$ in region~$r$ is scaled by $\phi_{grt}$.
If $\phi_{grt} = \phi_g^{(G)} \phi_{r}^{(R)}$, 
the group-specific effects $\phi_g^{(G)}$ will adjust the
columns of the contact matrix.

There are two special cases of the contact structure involved in the epidemic component.
First, a contact matrix with identical rows implies that the mixing pattern of
the $Y_{g',r',t-1}$ infectious cases does not depend on the group $g'$ they
belong to. An example of such homogeneous mixing is a contact matrix where each row
equals the vector of group sizes ($c_{g'g} = e_g$).
If $\phi_{grt}$ contains group-specific effects, 
a simple matrix of ones ($\bm{C} = \bm{1}$) will induce the same contact structure.
The other special case is a diagonal contact matrix $\bm{C} = \bm{I}$,
which reflects complete absence of mixing.
This is equivalent to formulating a separate spatio-temporal
model~\eqref{eqn:hhh4_twin} for each group.
However, also in this case of no between-group mixing, the joint model
formulation has the advantage of allowing for parsimonious decompositions
of $\nu_{grt}$ and $\phi_{grt}$ into group and region effects.
Borrowing strength across groups
is especially useful in applications with low counts.

\subsection{Parameterising the contact matrix}

Contact patterns derived from sociological studies might not fully match the
characteristics of disease spread. For example, social networks are known to
change during illness \citep{vankerckhove.etal2013} and brief contacts are
frequently not reported \citep{smieszek.etal2014}.
We therefore suggest a parsimonious single-parameter approach to adaptively
estimate the transmission weights as a function of the given contact
matrix~$\bm{C}$.

Our proposal is borrowed from 
\citet{kuechenhoff.etal2006}, who progressively transform
a misclassification matrix to establish an association between the amount of
misclassification in a covariate and the corresponding parameter estimate.
The proposed 
transformation is based on the eigendecomposition of the matrix $\bm{C}$
to raise it to the power of $\kappa \ge 0$,
\begin{equation} \label{eqn:powerC}
\bm{C}^\kappa := \bm{E} \bm{\Lambda}^\kappa \bm{E}^{-1} \:,
\end{equation}
where $\bm{\Lambda}$ is the diagonal matrix of eigenvalues and
$\bm{E}$ is the corresponding matrix of eigenvectors.  Translated to
our setting, the parameter $\kappa$ measures the amount of
transmission between the subgroups of the population.
Specifically, $\kappa = 0$ corresponds to complete absence of
between-group transmission ($\bm{C} = \bm{I}$), whereas $\kappa = 1$
leaves the contact matrix unchanged. If $\bm{C}$ is
row-normalized, 
all rows of $\bm{C}^\kappa$ converge to the same distribution
as $\kappa \to \infty$. 
The transmission pattern thus
becomes independent of the group the infected individual belongs to.
Because of this useful interpretation, 
we assume a pre-normalized
contact matrix $\bm{C}$ in the remainder of this paper.

\begin{figure}
\centering
\subfloat[$\bm{C}^\kappa$ for different values of $\kappa$.
\label{fig:powerC:matrix}]{

{\centering \includegraphics[width=0.6\linewidth]{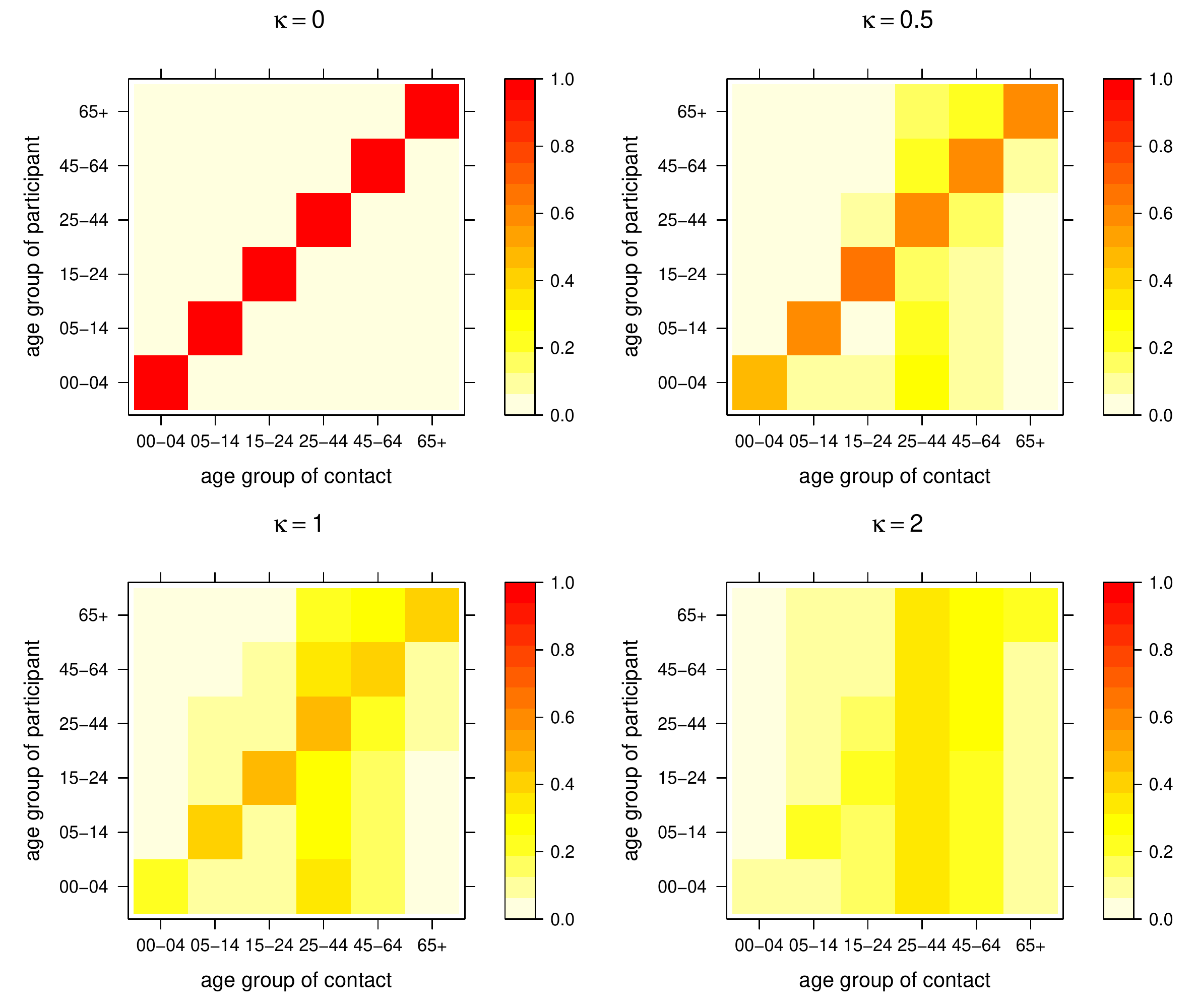} 

}

}
\subfloat[The diagonal entry $\lbrack 3,3\rbrack$ and the off-diagonal
entry $\lbrack 3,6\rbrack$ of $\bm{C}^\kappa$. 
\label{fig:powerC:entries}]{

{\centering \includegraphics[width=0.33\linewidth]{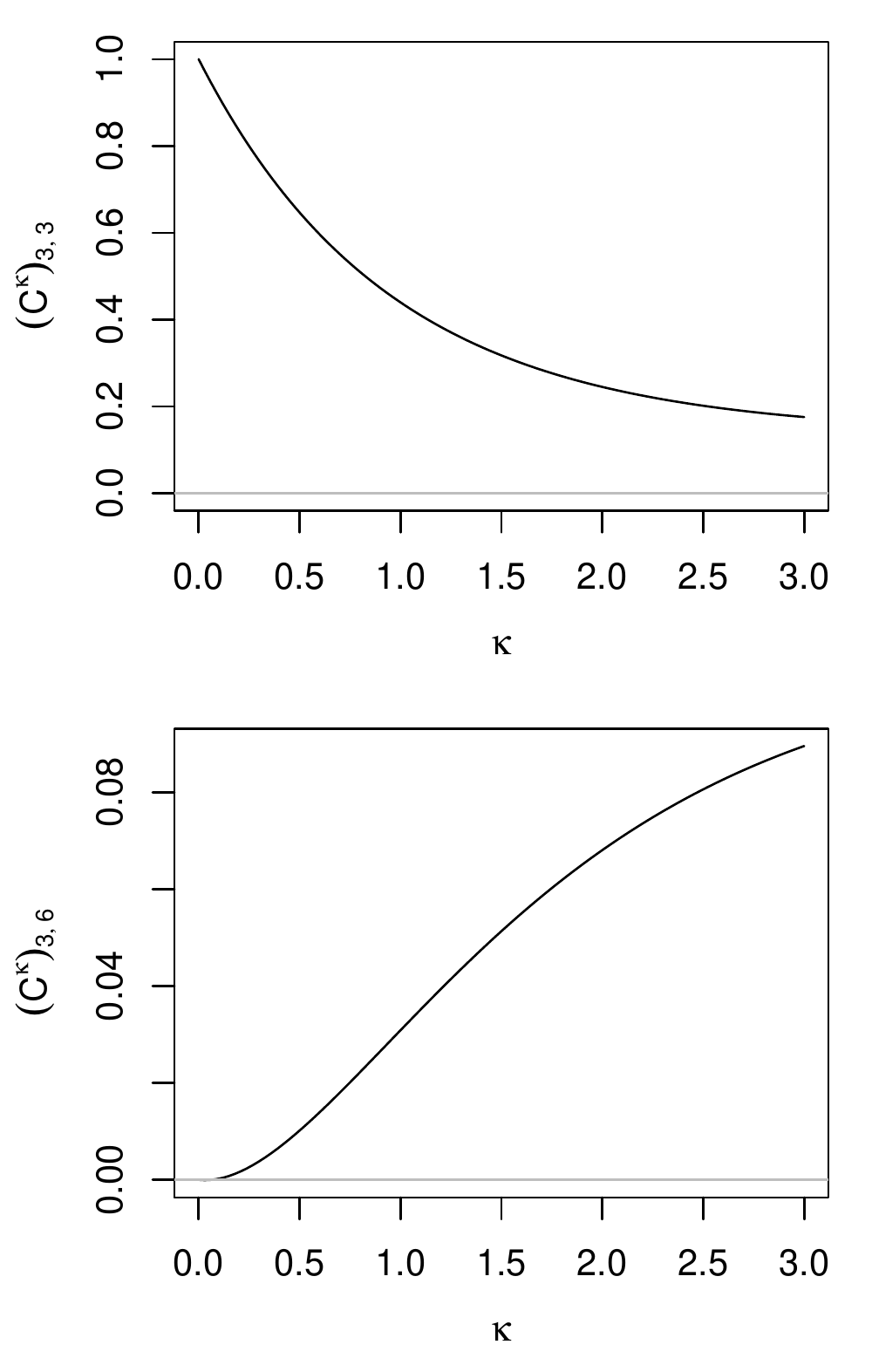} 

}

}
\caption{The power transformation \eqref{eqn:powerC} applied to the
  row-normalized POLYMOD contact matrix.}
\label{fig:powerC}
\end{figure}

The basic requirement
that $\bm{C}$ can be factorized by an eigendecomposition will hold in
most practical cases.  However, we also need to make sure that
$\bm{C}^\kappa$ has non-negative entries for $\kappa < 1$.  With our
contact matrix, two entries in $\bm{C}^\kappa$ become negative (but close to 0)
for small $\kappa$.  We follow a pragmatic
approach and truncate negative entries at 0.
Figure~\ref{fig:powerC} exemplifies $\bm{C}^\kappa$ for the
row-normalized version of the contact matrix from Figure~\ref{fig:Cgrouped},
and illustrates how diagonal and off-diagonal entries, respectively,
are affected by the power transformation.

\subsection{Inference}

Likelihood inference for the multivariate count time-series
model~\eqref{eqn:hhh4} has been developed by \citet{paul.held2011}
and \citet{meyer.held2013}.
The log-likelihood is maximized numerically using the quasi-Newton algorithm
provided by the \proglang{R}~function \code{nlminb} \citep{R:base}.
Supplied with analytical formulae for the score function and Fisher information,
convergence is fast, even for a large number of parameters.
The modelling framework is implemented in the \proglang{R}~package
\pkg{surveillance} \citep[Section~5]{meyer.etal2014} as function \code{hhh4}.

The age-structured model~\eqref{eqn:hhh4contacts} is built on
top of the existing inference framework. 
The power parameter $\kappa$ of \eqref{eqn:powerC} is
conveniently estimated via a profile likelihood approach
\citep[see, e.g.][Section 5.3]{Held.SabanesBove2014},
which avoids the cumbersome implementation of additional derivatives with
respect to all model parameters. We numerically maximize the
log-likelihood of a model with fixed contact matrix $\bm{C}^\kappa$ as a
function of $\kappa$. The profile confidence interval for
$\kappa$ thus incorporates the uncertainty of all other
parameter estimates (but not vice versa).

\section{Results} \label{sec:results}

We apply an age-structured spatio-temporal model
of the form~\eqref{eqn:hhh4contacts} to the norovirus
data described in Section~\ref{sec:data}.
As the number of cases varies strongly by age group,
we use group-specific overdispersion parameters $\psi_g$.
For the mean, we assume the endemic-epidemic structure
\begin{equation} \label{eqn:hhh4contacts_noroBE}
\begin{split}
  \mu_{grt} &= e_{gr} \, \exp\left\{ \alpha_g^{(G)} + \alpha_r^{(R)} +
    \beta x_t + \gamma_g \sin(\omega t) + \delta_g \cos(\omega t) \right\} \\
  &\quad + e_{gr}^\tau \, \phi_g^{(G)} \phi_r^{(R)} \sum_{g',r'}
  \normalize{(\bm{C}^\kappa)_{g'g} \, (o_{r'r} + 1)^{-\rho}} \, Y_{g',r',t-1} \:.
\end{split}
\end{equation}
The endemic predictor allows for age- and district-specific incidence levels,
fewer cases during the Christmas break ($x_t = 1$ in calendar
weeks 52 and 1, otherwise $x_t=0$),
as well as age-specific seasonality ($\omega = 2\pi/52$).
Transmission between age groups is modelled using the power transformation
\eqref{eqn:powerC} for the row-normalized
contact matrix estimated from the POLYMOD study.
Transmission between districts is quantified by a power law with respect to
adjacency order.
The intercepts are identifiable
by fixing $\alpha_1^{(G)}=\alpha_1^{(R)}=0$, $\phi_1^{(G)}=\phi_1^{(R)}=1$,
where $\phi_g^{(G)}$ and $\phi_r^{(R)}$ are estimated on the log-scale,
and including overall intercepts in both model components.


\begin{table}[p]
\centering
\caption{Model summaries for the age-stratified, areal surveillance data of norovirus gastroenteritis in Berlin. For reference, the first row represents the purely endemic model, which assumes independent counts. The remaining rows correspond to endemic-epidemic models with a spatial power law, but varying assumptions on the age-structured contact matrix $\bm{C}$. The columns refer to the following model characteristics: the number of parameters, the difference in Akaike's Information Criterion compared to the purely endemic model, the power $\tau$ of the population scaling factor, the decay parameter $\rho$ of the spatial power law, and the power adjustment $\kappa$ of the contact matrix. The parameter columns contain the estimates and 95\% Wald confidence intervals.} 
\label{tab:models}
\begingroup\small
\begin{tabular}{rlllll}
  \toprule
 & dim & $\Delta$AIC & $\tau$ & $\rho$ & $\kappa$ \\ 
  \midrule
purely endemic model & 36 & 0.0 & --- & --- & --- \\ 
   \midrule
homogeneous mixing ($\bm{C} = \bm{1}$) & 55 & -415.4 & 1.19 (0.83--1.55) & 2.43 (2.04--2.88) & --- \\ 
  no mixing ($\bm{C} = \bm{I}$) & 55 & -602.8 & 0.61 (0.24--0.98) & 2.18 (1.89--2.53) & --- \\ 
  original contact matrix $\bm{C}$ & 55 & -631.9 & 0.97 (0.66--1.28) & 2.34 (2.03--2.70) & --- \\ 
  adjusted contact matrix $\bm{C}^\kappa$ & 56 & -659.4 & 0.86 (0.53--1.19) & 2.27 (1.98--2.61) & 0.47 (0.34--0.66) \\ 
  based on physical contacts only & 56 & -655.3 & 0.85 (0.52--1.19) & 2.27 (1.98--2.61) & 0.48 (0.35--0.66) \\ 
   \bottomrule
\end{tabular}
\endgroup
\end{table}

Table~\ref{tab:models} summarizes competing models with respect to the assumed contact structure between age groups.
It turns out that a superposed epidemic component improves upon a
purely endemic model, and that incorporating the contact matrix from the POLYMOD
study outperforms naive models with homogeneous or no mixing between age groups.
Akaike's Information Criterion (AIC) is minimal for the model with a
power-adjusted contact matrix $\bm{C}^\kappa$ (penultimate row), where the exponent is
estimated to be $\hat{\kappa} =$
0.47 (95\% CI: 0.34 to 0.66).
This means that the epidemic part subsumes more information from cases in the
own age group than suggested by the original contact matrix
(\cf Figure~\ref{fig:powerC:matrix}).
The change in AIC associated with this adjustment, however, is minor
compared to the improvement achieved by
employing the POLYMOD contact matrix in the first place.
Results are very similar for \emph{physical} contacts, but the fit is
slightly worse.

The spatial spread of the disease across city districts is estimated to have a
strong distance decay with $\hat{\rho} =$
2.27 (95\% CI: 1.98 to 2.61),
such that the adjacency orders 0 to 4 have weights
1.00, 0.21, 0.08, 0.04, and 0.03.
Supplementary Figure~S4 shows age-dependent power laws
(replacing $\rho$ by $\rho_{g'}$ in~\eqref{eqn:hhh4contacts_noroBE}),
as well as unconstrained
estimates of the order-specific weights, 
which are close to the power law.
In accordance with the idea of a gravity model, 
we find that the
epidemic part scales with the population size of the ``importing'' district and
age group. 
Similar to a previous application on influenza \citep{meyer.held2013}, the
corresponding estimate $\hat{\tau} =$
0.86 (95\% CI: 0.53 to 1.19)
is slightly below unity and provides strong evidence for such an association.

\begin{figure}

{\centering \includegraphics[width=\linewidth]{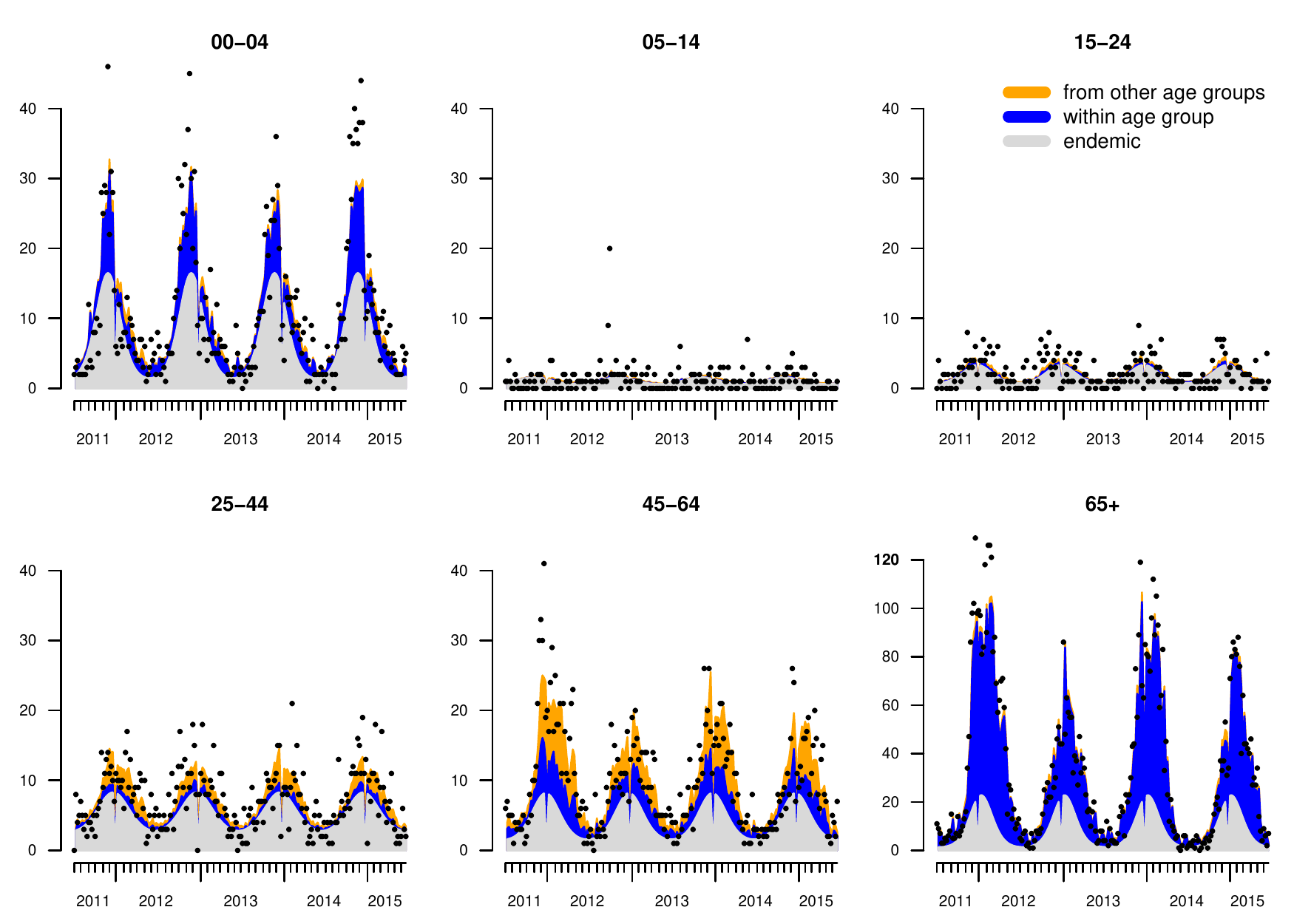} 

}

\caption{Fitted mean components from the AIC-optimal model with adjusted contact matrix, aggregated over all districts. The dots correspond to the reported numbers of cases.}\label{fig:fitted_groups}
\end{figure}

Figure~\ref{fig:fitted_groups}
shows the endemic-epidemic decomposition of the estimated mean aggregated across
districts (see the supplementary Figures~S5 to S7 for the district-level and
overall fits).
When reformulating the
model as a multivariate branching process with immigration
\citep{held.paul2012}, the largest eigenvalue of the 
matrix holding the estimated coefficients of $Y_{g',r',t-1}$ is
0.71,
which can be interpreted as the overall epidemic proportion of disease incidence.
However, this value mostly reflects the situation for the
65+ age group where the within-group spread is dominating.
In contrast, for the groups of 5--14 and 15--24 year-old persons, almost no
dependence on past counts of the same or the other age groups can be identified.
Interestingly, the groups of 25--44 and 45--64 year-old persons seem to
inherit a relevant proportion of cases from other age groups.
The youngest age group, though, mostly depends on the endemic component and its
own cases,
which is probably related to their early onset.
The age-dependent sine-cosine effects 
capture these shifts and are shown in supplementary Figure~S8.
The modal endemic incidence is in calendar weeks
48 (0--4), 45 (5--14), 52 (15--24), 51 (25--44), 52 (45--64), and 3 (65+),
respectively. The largest amplitude is estimated for the youngest and oldest
groups.

The estimated group-specific overdispersion parameters are
0.24 (0--4), 1.98 (5--14), 0.30 (15--24), 0.03 (25--44), 0.15 (45--64), and 0.40 (65+)
in the model with adjusted contact matrix.
The large overdispersion for the 5- to 14-year-old children may be partly due to
the food-borne outbreak
in 2012, for which the model does not explicitly account.
The estimates are similar for the other
epidemic models of Table~\ref{tab:models}, but slightly larger in the
endemic-only model.


\section{Discussion}
\label{sec:discussion}

We have incorporated a social contact matrix
in a regression-oriented, endemic-epidemic time-series model for
stratified, area-level infectious disease counts.
This three-dimensional approach provides a more detailed description
of disease spread than unstratified or non-spatial models, which
inherently assume homogeneous mixing within each region or subgroup, respectively.

In our application to age-stratified counts of norovirus gastroenteritis in Berlin's city districts, 
the contact model was superior to 
homogeneous or no mixing between age groups.
The model further improved when adjusting the POLYMOD contact matrix
towards more within-group transmission.
This could be related to biases in contact reporting \citep{smieszek.etal2014}
with more unreported (short) contacts along the diagonal.
The two age groups involving parents were affected the most by
preceding infections in other age groups. This is in accordance with
the leading role of school children 
in influenza epidemics \citep{worby.etal2015}. 

Furthermore, new infections predominantly depend on past cases from the same
district, 
as suggested by the estimated spatial transmission weights.
An age-dependent distance decay could not be identified from the disease counts.
One could thus try to replace the parametric formulation by a social
contact matrix, stratified by spatial distance in addition to age group.
Separate movement data for school children and adults could then be used to quantify
the strength of epidemiological coupling between regions
\citep{kucharski.etal2015}. However, integration of movement network data
does not necessarily improve predictions \citep{geilhufe.etal2012}.

A potentially more severe simplification of our model is the assumption
of a time-constant contact matrix. Although weekday vs.\ weekend
differences in contact patterns are not relevant for weekly time-series models, 
there are possibly relevant seasonal effects on larger time scales. For instance, the
contact structure of school children changes considerably between regular and
school holiday periods \citep{hens.etal2009}.
Our model could be further tuned both by incorporating a time-varying contact matrix and
by estimating seasonality also in the epidemic component \citep{held.paul2012},
which the \code{hhh4} implementation already supports.

To check the robustness of our results with respect to under-reporting,
we re-estimated the models with age-specific multiplication factors applied to
the reported numbers of cases.
Roughly following \citet[Table~1]{bernard.etal2014}, we used factors of
1.5 (0--4), 2.5 (5--14), 3.0 (15--24), 3.0 (25--44), 2.5 (45--64), and 2.0 (65+),
respectively. While the overdispersion increases, the parameters of the mean
are close to the original fit and the epidemic proportion
is similar (supplementary Figure~S9).
For small strata with a low number of cases, a drawback of this simple
deterministic approach is that zero reported counts remain zero regardless of
the amount of under-reporting. More sophisticated adjustments
are currently being investigated within a Bayesian modelling 
framework. In principle, asymptomatic infections could be similarly accounted
for as missing cases, but they seem to play a minor role in disease
transmission \citep{sukhrie.etal2012}. 
One-week-ahead forecasts or long-term simulations of the number
of (symptomatic) infections, however, are of particular relevance for public health planning.
Whether the improved model with social contact data also leads to better
predictions will be described elsewhere.

\section*{Acknowledgements}

We thank the associate editor, two anonymous referees, and Michael H\"ohle
for helpful comments on a previous version of this manuscript.
Jo{\"e}l Mossong made the POLYMOD data available, and
a KML file of Berlin's districts 
was obtained from the Statistical Office of Berlin-Brandenburg.

\section*{Funding}

Swiss National Science Foundation (project \#137919).

\section*{Supplementary material}

Supplementary material is available at \url{http://biostatistics.oxfordjournals.org}.



\renewcommand{\bibfont}{\small}
\setlength{\bibsep}{4pt plus .3ex}
  \bibliographystyle{biorefs}
  \bibliography{references}

\end{document}